\documentclass[aps,prx,preprint,superscriptaddress]{revtex4-1}
\usepackage{graphicx}
\usepackage{epstopdf}

\begin{document}

\title{Surface and bulk superconductivity at ambient pressure in the Weyl semimetal TaP }

\author{M. R. van Delft}
\affiliation{High Field Magnet Laboratory (HFML-EMFL), Radboud University, Toernooiveld 7, 6525 ED Nijmegen, The Netherlands }
\affiliation{Radboud University, Institute for Molecules and Materials, Nijmegen 6525 AJ, The Netherlands}
\author{ S. Pezzini}
\affiliation{High Field Magnet Laboratory (HFML-EMFL), Radboud University, Toernooiveld 7, 6525 ED Nijmegen, The Netherlands }
\affiliation{Radboud University, Institute for Molecules and Materials, Nijmegen 6525 AJ, The Netherlands}
\author{M. K{\"o}nig}
\affiliation{ Max Planck Institute for Chemical Physics of Solids, D-01187 Dresden, Germany}
\author{ P. Tinnemans}
\affiliation{Radboud University, Institute for Molecules and Materials, Nijmegen 6525 AJ, The Netherlands}
\author{ N. E. Hussey}
\affiliation{High Field Magnet Laboratory (HFML-EMFL), Radboud University, Toernooiveld 7, 6525 ED Nijmegen, The Netherlands }
\affiliation{Radboud University, Institute for Molecules and Materials, Nijmegen 6525 AJ, The Netherlands}
\author{ S. Wiedmann}
\affiliation{High Field Magnet Laboratory (HFML-EMFL), Radboud University, Toernooiveld 7, 6525 ED Nijmegen, The Netherlands }
\affiliation{Radboud University, Institute for Molecules and Materials, Nijmegen 6525 AJ, The Netherlands}

\begin{abstract}
The motivation to search for signatures of superconductivity in Weyl semi-metals and other topological phases lies in their potential for hosting exotic phenomena such as nonzero-momentum pairing or the Majorana fermion, a viable candidate for the ultimate realization of a scalable quantum computer. Until now, however, all known reports of superconductivity in Weyl semi-metals have arisen through surface contact with a sharp tip, focused ion-beam surface treatment or the application of high pressures. Here, we demonstrate the observation of superconductivity in single crystals, even an as-grown crystal, of the Weyl semi-metal tantalum phosphide (TaP), at ambient pressure. A superconducting transition temperature, $T_c$, varying between 1.7 and 5.3 K, is observed in different samples, both as-grown and microscopic samples processed with focused ion beam (FIB) etching. Our data show that the superconductivity present in the as-grown crystal is inhomogeneous yet exists in the bulk. For samples fabricated with FIB, we observe, in addition to the bulk superconductivity, a second superconducting state that resides on the sample surface. Through measurements of the characteristic fields as a function of temperature and angle, we are able to confirm the dimensionality of the two distinct superconducting phases.  
\end{abstract}

\maketitle

\section*{Introduction}
Since the discovery of Weyl semimetals, a great deal of work has been devoted to understanding the properties of these topological materials, whose band structure includes specific points known as Weyl nodes where non-degenerate bands touch each other and disperse linearly. Weyl semimetals differ from the related Dirac semimetals in that they require either time-reversal or inversion symmetry to be broken in order to lift the degeneracy of the nodes. Consequently, Weyl nodes always exist in pairs of opposite chirality that are connected through Fermi arcs running along the surface of the material, as has been observed experimentally in several materials using ARPES \cite{Deng2016,Lv2015a,Xu2015,Xu2015b,Xu2015e,Souma2015}. The chiral nature of the Weyl nodes can furthermore manifest itself in electrical transport, in the form of the chiral anomaly, leading to a negative longitudinal magnetoresistance. The observation of this effect has been reported in several materials \cite{Niemann2016,Huang2015a,Zhang2015d}, but its origin remains uncertain \cite{Reis2016b,Li2017a}.

In addition to the intrinsic transport properties of Weyl semimetals, the combination of Weyl physics and superconductivity may support Majorana \cite{Chen2016d} or other exotic surface states \cite{Lu2015a} as a result of their topological nature. These states are of fundamental interest and may eventually be applicable in the field of quantum computation. For this reason, there is an ongoing effort aimed at achieving superconductivity in such materials and investigating their properties, either through the use of the proximity effect \cite{Khanna2014,Chen2016d} or by other means.

The family of compounds comprising TaP, TaAs, NbAs and NbP are all experimentally confirmed as Weyl semimetals \cite{Xu2015d,Lv2015a,Huang2015d,Xu2015,Xu2015e,Shekhar2015a} and are now among the most ardently studied compounds in this class. Under the right conditions, each member of the family has shown trace signatures of superconductivity. In TaAs, a superconducting onset has been induced through contact with a sharp tip of either Ag \cite{Aggarwal2016b} or PtIr \cite{Wang2017a}, while in TaP, this onset was achieved under the application of extremely high pressures \cite{Li2017b}. In each case, however, only partial drops in the resistivity were observed. A state of zero resistance has only been reported in samples where superconductivity was induced in a thin surface layer by treatment with focused ion beam (FIB)  \cite{Bachmann2017}. To date, no trace of superconductivity at ambient pressure has been reported in pristine samples of any member of this family.

In this work, we demonstrate the existence of intrinsic bulk superconductivity in TaP, in addition to the FIB-induced surface superconductivity that was previously reported by Bachmann \emph{et al.} \cite{Bachmann2017}. In our FIB-processed crystals (an example of which is shown in Fig. \ref{Fig:TaP}a), both bulk and surface superconductivity are observed, while in the as-grown parent crystal only bulk but inhomogeneous superconductivity appears.  In both cases, we find $T_c$ to vary between 1.7 and 5.3 K. Finally, we delineate and identify the two transitions in the microfabricated crystals based on the angle dependence of their characteristic magnetic fields \cite{Li2017b}.

  \section*{Results}
\begin{figure} 
 \includegraphics [width=0.5\textwidth]{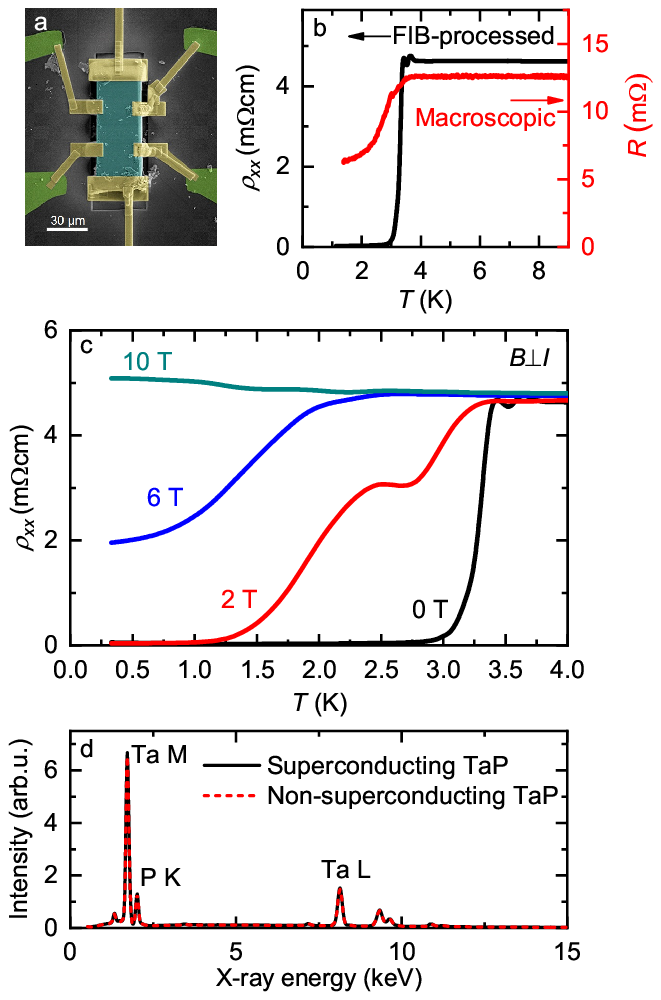} 
 \caption{\textbf{a} False color SEM image of a typical FIB contacted crystal.  \textbf{b} Temperature sweeps showing the resistive transition of the FIB-processed sample 4 and the macroscopic crystal from which all FIB samples were cut.  \textbf{c} Resistivity curves at low temperature of FIB-processed sample 4, under different, constant magnetic fields perpendicular to the current direction. \textbf{d} EDX spectra of the bulk sample used in our experiments and a non-superconducting sample obtained from a different source. Within the resolution, the compositions are the same.   
	\label{Fig:TaP}}
 \end{figure}

    Fig. \ref{Fig:TaP}b shows low-temperature resistivity curves for both the pristine crystal and one of the microstructured samples (data for all four microstructured samples can be found in Fig. S1 of the Supplementary Information). The parent crystal exhibits an incomplete, resistive transition at a transition temperature $T_c$=3$\pm$0.8 K that resembles the one which had been observed previously in TaP under high pressure \cite{Li2017b}. The microstructured crystal, on the other hand, exhibits a sharp transition to zero resistance with $T_c$=3.3$\pm$0.5 K.
  
  In order to confirm the presence of superconductivity in our samples, we studied the evolution of the resistive transition in the presence of a magnetic field. Fig. \ref{Fig:TaP}c shows temperature sweeps of the resistivity of one of the microstructured samples under different, constant magnetic fields. These fields are applied perpendicular to the current direction. Clearly, the resistive transition in the FIB-processed sample is suppressed gradually with field, as is expected in the case of a superconducting transition. For $B_{\perp}$=10 T, it vanishes completely. In the parent crystal, the transition is suppressed more strongly and vanishes under a very small magnetic field. 
  
Superconductivity has been reported before in microfabricated samples of TaP, TaAs, NbP and NbAs \cite{Bachmann2017}, but in as-grown TaP it has never been observed at ambient pressure despite the large number of measurements at low temperature that have been carried out on this material \cite{Arnold2016,Besara2016,Du2016,Hu2016}. For this reason, we investigate our bulk crystal for any deviations in terms of composition and structure that might lead to superconductivity.

  We studied the parent crystal with EDX and compared the stoichiometry with that of a crystal of TaP from a different source that did not display any sign of superconductivity. The resulting spectra are shown in Fig. \ref{Fig:TaP}d. From these we find that the two crystals have a stoichiometry that is identical to within an experimental error of about 1\%, proving that our material is very similar to that used in other studies. Furthermore, XRD measurements demonstrate that our TaP crystal is in the $I4_1 md$ space group, as is usual for TaP under ambient conditions \cite{Willerström1984,Xu2015d,Xu2016}. This is different, however, to that found in TaP under high pressure or in MoP (also under pressure), where in both cases superconductivity appears in the $P$-$6m2$ phase. Thus the superconducting state that develops in our crystals appears to be distinct from what has been observed before.
    
\begin{figure} 
 \includegraphics [width=0.5\textwidth]{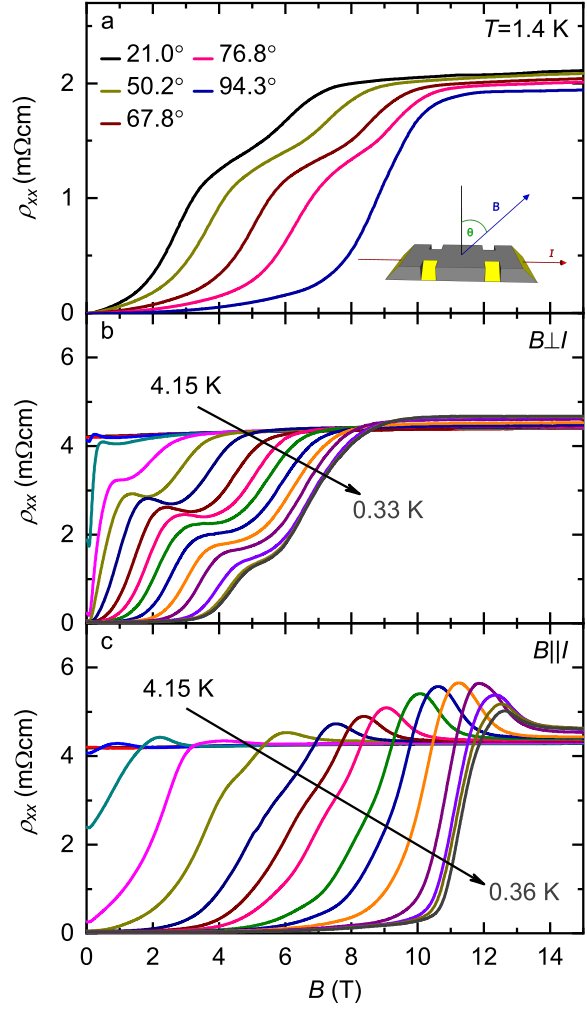} 
\caption{Field sweeps of the resistivity of the FIB-processed sample 4 for \textbf{a} different angles as defined in the inset and \textbf{b} different temperatures in a magnetic field parallel to the current direction and \textbf{c} in a field perpendicular to the current. \label{Fig:Fieldsweeps}}
 \end{figure}

 Fig. \ref{Fig:Fieldsweeps}a shows field sweeps of the resistivity of one of the microstructured crystals (sample 4) under different orientations of the magnetic field. Here, the angle 90$^{\circ}$ denotes a field parallel to the direction of the applied current (see inset of Fig. \ref{Fig:Fieldsweeps}b), which in this case is the crystallographic $ab$-plane. For angles away from  90$^{\circ}$, there is a clear double transition that we associate with distinct superconducting states. Close to  90$^{\circ}$, the characteristic fields of the two superconducting states appear to merge, causing the two transitions to become indistinguishable.
 
  Figs.  \ref{Fig:Fieldsweeps}b and c show the temperature dependence of the resistive transitions in field, for fields perpendicular and parallel to the current direction respectively. In perpendicular field (Fig. \ref{Fig:Fieldsweeps}b), there are two distinct superconducting features separated by a broad shoulder. In the parallel field configuration on the other hand, the shoulder is much weaker. This suggests that the characteristic fields of the two states are of similar magnitude, but have a different temperature dependence. The data also reveal an upturn of the resistivity before the normal state is fully restored. Such an upturn is frequently observed in inhomogeneous superconductors \cite{Jaroszynski2008,Santhanam1991,Nordstrom1992,Klimczuk2003} due to current redistribution as some parts of the crystal turn superconducting while others remain resistive \cite{Vaglio1993}. It was also recently seen in the pressure induced superconducting state of MoP \cite{Chi2018}.
  
\section*{Discussion}
\subsection*{Dimensionality analysis}

\begin{figure} 
 \includegraphics [width=0.5\textwidth]{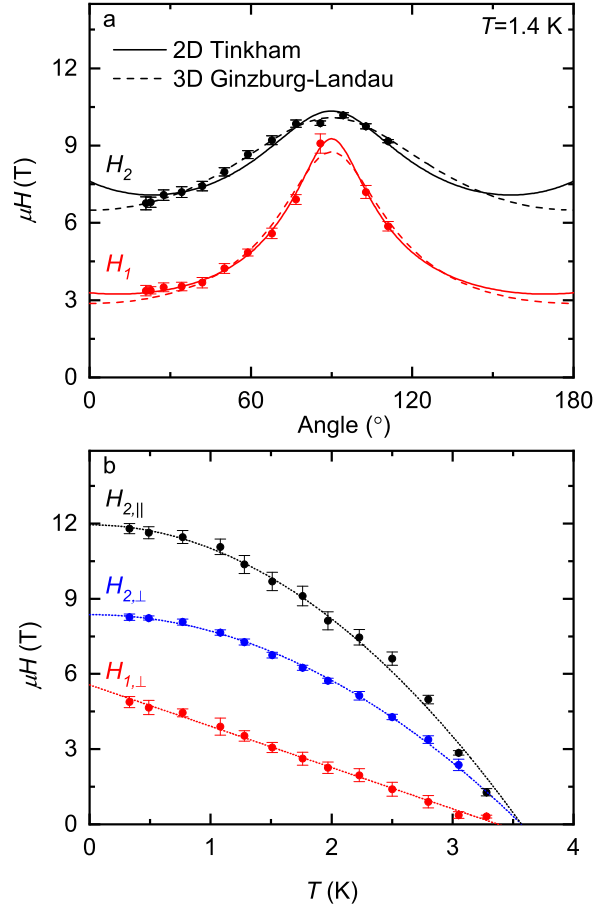} 
 \caption{\textbf{a} Angle dependence of the characteristic fields, fitted with the two-dimensional Tinkham model (solid lines) and the three-dimensional Ginzburg-Landau model (dashed lines). \textbf{b} Temperature dependence of the characteristic fields for parallel and perpendicular fields. $H_{2}$ is fitted using a three-dimensional model and $H_{1}$ with the two-dimensional Ginzburg-Landau model. $H_{1}$ in parallel field cannot be unambiguously identified and is thus omitted from this figure.\label{Fig:2D3D}}
 \end{figure}

 The data in Fig. \ref{Fig:Fieldsweeps} can be used to extract characteristic fields for both transitions as a function of angle and temperature, the results of which are shown in Fig. \ref{Fig:2D3D}. The two field scales are determined by defining the field at which the resistance has risen to 90\%  of the normal state ($H_2$) and the field that minimizes the second derivative of the resistivity ($H_1$) (see supplementary Fig. S2 for more detail). In Fig.  \ref{Fig:2D3D}a, we show both $H_1$ and  $H_{2}$ as a function of the angle between the magnetic field and the $ab$-plane. The solid and dashed lines are fits to the two-dimensional Tinkham model \cite{Tinkham1963} and the three-dimensional Ginzburg-Landau (GL) model \cite{Tinkham2012}, respectively. It is found that the 2D Tinkham model gives an excellent description of the behavior of $H_{1}$, covering not only the cusp at 90$^{\circ}$ which the 3D GL misses, but also its behavior in near perpendicular fields. Conversely, the angle dependence of $H_{2}$ does not have a sharp cusp as $H_{1}$ does and is better described by the 3D GL model. 
 
 Some refinements of the 2D Tinkham model are possible (outlined in the Supplementary Information), one for the case of intrinsic surface superconductivity and another for a thin superconducting film with a thickness less than or comparable to the coherence length $\xi_{GL}$, a less stringent requirement than that of the Tinkham model which requires a thickness $d\ll\xi_{GL}$ \cite{Yamafuji1966,Yamafuji1966a}. Both of these adapted models, however, lead to a less accurate fitting for both $H_{1}$ and $H_{2}$ (see supplementary Fig. S3). Thus, with the GL and Tinkham models giving the most accurate descriptions of $H_{2}$ and  $H_{1}$ respectively, we conclude that $H_{2}$ arises from the bulk crystal while $H_{1}$ is characteristic of a 2D superconducting state, presumably arising from a very thin layer on the surface that satisfies the criterion of $d\ll\xi_{GL}$.
 
 Fig. \ref{Fig:2D3D}b shows the phase diagram of FIB-processed TaP, in both parallel and perpendicular field configurations. $H_{1}$  is only shown for perpendicular field, as it is not distinguishable in the parallel field configuration. 
The behavior of $H_{2}$ in both configurations can be well described by the expression $H_2=H_2(0)(1-(T/T_c)^2)$, in agreement with the description of the critical field of a three-dimensional GL superconductor.  $H_{1}$ on the other hand, is better described with a linear temperature dependence, consistent with the GL model for two-dimensional superconductors: $\mu H_{1,\perp}=\Phi_0/(2\pi \xi_{GL}^2)(1-T/T_c)$ \cite{Lu2015}. Associating $H_1$ with the upper critical field for the two-dimensional superconducting layer, we can estimate $\xi_{GL}$ to be approximately 7.7 nm.

In order to make an estimate of the thickness $d$ of the superconducting layer, we use the SRIM-2013 \cite{Ziegler2013} code to simulate the ion milling process. With the low acceleration voltage of 8 kV that is used for the final polishing of the sample, the Ga$^+$ ions penetrate about 5.0 nm below the surface of the sample. As P is sputtered approximately two times more readily than Ta, an average composition is expected in this thin layer of  Ta$_{2.1}$P. Of course, the thickness of the superconducting layer cannot simply be assumed to be the same as the ion penetration depth; if anything, this depth gives an upper limit. Considering this, it is not unreasonable to expect that $d\ll\xi_{GL}$ is indeed satisfied.

\subsection*{Critical currents}
Information on the evolution of the two superconducting phases with temperature can be gleaned by looking at the critical currents associated with the transitions. To this end, we measured the current-voltage (IV) characteristics and differential resistivity curves in different magnetic fields and temperatures. In Fig. \ref{Fig:dVdI}, the data are shown for zero field at $T$=1.3 K (see supplementary Figs. S4 and S5 for the full set of data in different magnetic fields and at different temperatures).

Several clear transitions can be seen in the differential curves shown in Fig. \ref{Fig:dVdI}b. Around 1.0 and 1.1 mA, there are two distinct features that do not lead to a significant change in resistance. We presume that these correspond to parts of the bulk crystal becoming superconducting while others remain resistive and are likely to be a consequence of strong inhomogeneity present in the sample. The transition at 0.7 mA then represents the majority of the bulk crystal becoming normal, leading to a strong increase in the resistivity. The state of zero resistance, however, can only be seen below a much smaller bias current of about 60 $\mu$A, as seen in the insets of Figs. \ref{Fig:dVdI}a and b. Considering the two-dimensional nature of the surface layer superconductivity, this small bias current corresponds to a current density of approximately 8$\times 10^4$ A$\cdot$cm$^{-2}$, much larger than that associated with the bulk transition (about 1$\times 10^3$ A$\cdot$cm$^{-2}$).

Figs. \ref{Fig:dVdI}c-e show the critical currents associated with the bulk and surface superconducting phases of the FIB-processed TaP as a function of temperature and magnetic field. This is a further confirmation of Fig. \ref{Fig:2D3D} as the two transitions are suppressed at the same temperature, but at different values of the magnetic field. In perpendicular field, the feature associated with surface superconductivity can be seen up to 1 T, whereas in parallel field it survives up to 7 T. The difference is less apparent for the bulk superconductivity, but also this can be seen to persist to higher fields in the parallel than in the perpendicular configuration, in agreement with Fig. \ref{Fig:2D3D}b.

\subsection*{BCS-BEC crossover regime}
Via Hall effect measurements (one of which is shown in supplementary Fig. S6a), the carrier concentration $n$ can be extracted for each of the microstructured samples, giving values of 4.5$\times 10^{18}$ to 3.9$\times 10^{19}$  cm$^{-3}$. For a semimetal such as TaP, these are typical values, in agreement with the literature \cite{Arnold2016,Du2016}. However, for a superconducting material, these are unusually low carrier concentrations. For comparison, we consider SrTiO$_3$, for which a carrier concentration of 4.1$\times 10^{18}$ with a $T_c$ of 180 mK was reported \cite{Lin2013}. The $T_c$ of our TaP is at least 10 times higher with a comparable carrier density. For these reasons, we consider the possibility of a crossover between a Bardeen-Cooper-Schrieffer state and a Bose-Einstein condensate (BCS-BEC) for the observed superconductivity. 

Using the value of 7.7 nm obtained from the fitting of Fig. \ref{Fig:2D3D} for the coherence length $\xi$, we calculate the number of pairs in the coherence volume $V_{coh}=4/3 \pi \xi^3$. In sample 4, this amounts to approximately 25 pairs, suggesting there is limited overlap between the pairs. Typically, a BCS superconductor has many thousands of pairs overlapping in $V_{coh}$, whereas a BEC superconductor has less than one pair in $V_{coh}$ and there is no interaction between different pairs. With a number of 25 pairs in $V_{coh}$, sample 4 is similar to FeSe (with 31 pairs \cite{Yang2017}), which is considered to be in a BCS-BEC crossover regime \cite{Kasahara2014,Kasahara2016,Watashige2017}. These findings therefore suggest that this new breed of semi-metals is a good playground for the observation of possible exotic superconductivity on the BCS-BEC boundary.

\section*{Conclusions}
In conclusion, we have established the existence of bulk, inhomogeneous superconductivity at ambient pressure in a crystal of TaP and confirmed the appearance of FIB-induced surface superconductivity. In other studies, it was found that TaP typically contains a large density of defects and can be off-stoichiometric with an excess of Ta \cite{Besara2016,Willerström1984}. Our EDX data do not exclude off-stoichiometry in our samples, but considering the similarity between the superconducting and non-superconducting samples, any overall off-stoichiometry cannot explain the superconductivity we observe. 

   It is, however, apparent that our sample is strongly inhomogeneous as we see multiple partial superconducting transitions, as well as an upturn of the resistance just above critical field in the parallel field configuration. As such there may exist domains with a local excess of Ta or defect structures that support superconductivity. Nevertheless, our findings call for a thorough study of the growth of TaP and related compounds in order to establish under what conditions superconductivity is optimized. Further research is also required to ascertain whether bulk, superconducting TaP retains all the characteristics of a Weyl semimetal. If it does, TaP may provide an ideal platform for the study of Weyl superconductivity.

\section*{Methods} 
\subsection*{Sample preparation}
The single crystal used in this study was grown by chemical vapor transport using polycrystalline TaP as a source material \cite{Xu2016}. Via X-ray crystallography, a facet matching the ab-plane of this crystal was identified and the microsamples were cut from this facet using focused ion beam (FIB) milling. For the rough cutting, an acceleration voltage of 30 kV with a large current of 20 nA was used. Initially, a rectangular piece of the crystal was cut out and remained attached only via a thin bar to the main crystal. A micromanipulator was then brought in contact with the rectangle and fixed onto it via Pt deposition, after which the bar was cut through and the sample was transferred to a silicon oxide substrate with prepatterned gold contact pads. On the substrate, a further shaping of the sample took place at 30 kV and 0.9 nA, followed by a more precise cleaning at 8 kV and 0.2 nA. Contacts were made between the sample and the gold pads via FIB-induced Pt deposition in a standard Hall-bar configuration (see Fig. \ref{Fig:TaP}a).

\subsection*{Resistivity measurements}
The resistivity measurements were performed in either a superconducting magnet with a maximum field of 15 T, using a $^3$He cryostat, or a resistive magnet of 33 T with a $^4$He cryostat with base temperature 1.3 K. We mounted the samples on a rotatable platform in order to vary their angle with the field. Each microsample had six contacts, with the two covering the short ends of the sample being used as current contacts and the other contacts to measure either a longitudinal or a hall voltage. We used an AC current excitation of 10 or 100 $\mu$A and acquisition took place with standard lock-in techniques. For the measurement of differential resistance, we used a voltage source in series with a 100 k$\Omega$ resistor to supply the DC bias current, with the lock-in amplifier similarly supplying a 10 $\mu$A AC current on top. A multimeter was placed in parallel with the lock-in amplifier to measure the DC signal. 

The as-grown sample had an arbitrary shape and so it could not be contacted in any well-defined geometry, making it impossible to determine the resistivity. This sample was measured with four contacts placed along the crystal and a current of 0.5 mA.

\subsection*{Energy-dispersive X-ray analysis (EDX)}
In order to determine whether off-stoichiometry might be responsible for the observed superconductivity in our samples, we performed our elemental analysis of the parent crystal together with a non-superconducting crystal of TaP from a different source. The two bulk crystals were measured in the same EDX system during a single run to exclude any difference in signal other than from the material itself. We aligned the crystals by eye to have a flat surface facing the electron beam. To correct for any small discrepancies remaining in the angle between the electron beam and the crystal, both crystals were measured twice with a 180$^{\circ}$ in-plane rotation in between and the two measurements were averaged. The results before and after the rotation were comparable, suggesting that the orientation and flatness of the surfaces were good. 
      
Despite our best efforts, we could not determine the precise Ta:P ratio with real confidence due to the inherent difficulties associated with quantitative EDX measurements. Such a measurement would require a reference TaP sample that is precisely stoichiometric. While our reference sample is known to be non-superconducting, its stoichiometry is not guaranteed. 

\subsection*{X-Ray Diffraction (XRD)}
Reflections were measured on a Bruker D8 Quest diffractometer with sealed tube and Triumph monochromator ($\lambda$ = 0.71073 \AA).  The unit cell was found using the software CELL\_ NOW \cite{Sheldrick2008}. 

\subsection*{Data availability}
The data that support the findings of this study are available from the corresponding author upon request.

\section*{Acknowledgments}
The authors would like to thank Nan Xu for providing the sample used to conduct this study and Andrew Mackenzie for help with the interpretation of the EDX data. We also acknowledge the support of the HFML-RU/FOM, member of the European Magnetic Field Laboratory (EMFL).

\section*{Author contributions}
M.R.v.D. processed the samples with FIB. M.R.v.D. and S.P. performed the electrical transport measurements. M.R.v.D., S.P., N.E.H. and S.W. participated in the interpretation of the transport data. M.K. performed the EDX measurements and P.T. the XRD measurements. M.R.v.D. analyzed the data and prepared the manuscript, which was completed with input from all authors.

\section*{Additional information} 

\textbf{Competing interests:} The authors declare no competing interests.

\bibliographystyle{unsrt}
\bibliography{library}

  \begin{figure} 
 \includegraphics [width=1\textwidth]{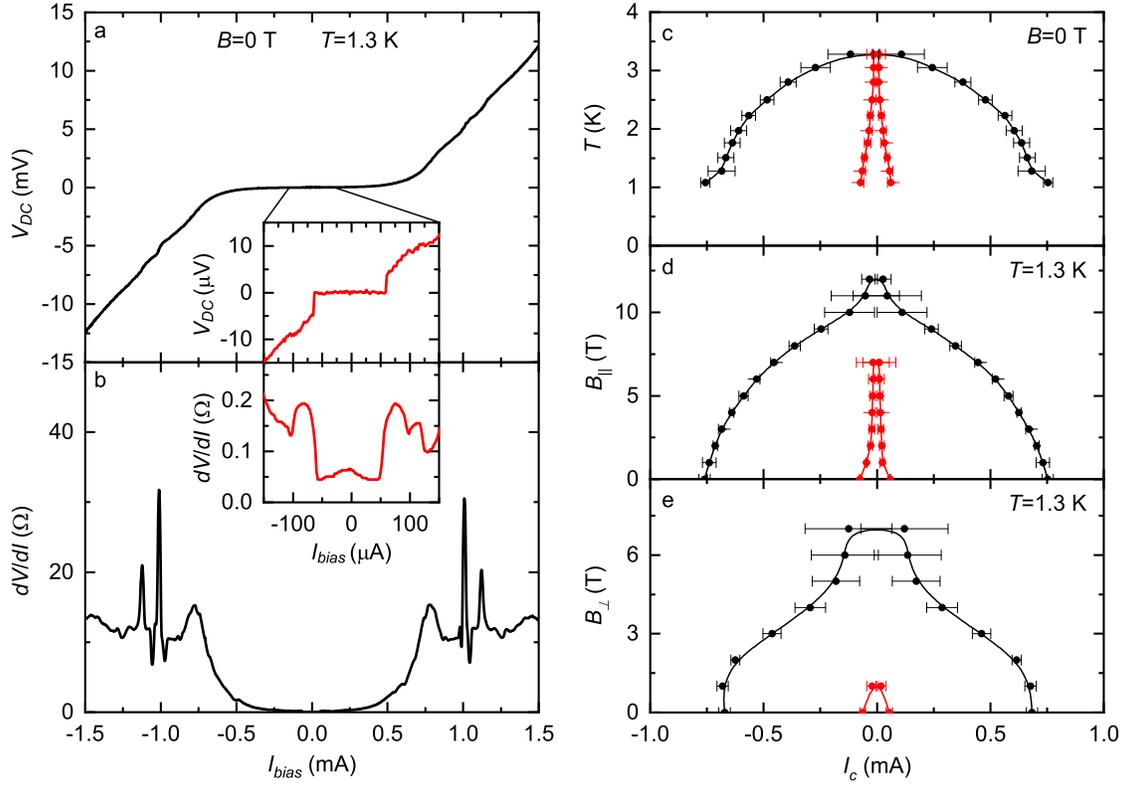} 
\caption{ \textbf{a} IV curve of sample measured at $T$=1.3 K and $B$=0 T. Inset: expanded view on the low-current region where the transition due to the surface can be seen. \textbf{b} Differential resistance measured simultaneously with the IV curve. Inset: differential resistance corresponding to the inset of \textbf{a}. \textbf{c}-\textbf{e} Critical currents of the bulk (black) and surface (red) as a function of \textbf{c} the temperature, \textbf{d} parallel magnetic field and \textbf{e} perpendicular magnetic field.
	\label{Fig:dVdI}}
 \end{figure}
\end{document}